\documentclass[twocolumn,floats,floatfix,superscriptaddress,prl,nofootinbib]{revtex4}
\usepackage{bm}
\usepackage{float}
\usepackage{graphicx}
\usepackage{amsmath}
\usepackage{amsfonts}
\usepackage{mathrsfs}
\usepackage{bbold}

\usepackage[utf8]{inputenc}
\usepackage{tabularx}

\usepackage{color}

\usepackage{color}

\begin{document}

\title{Making the Cut:  Lattice \textsl{Kirigami} Rules}

\author{Toen Castle}
\affiliation{Department of Physics and Astronomy, University of Pennsylvania, 209 South 33rd Street, Philadelphia, Pennsylvania 19104, USA}
\author{Yigil Cho}
\affiliation{Department of Physics and Astronomy, University of Pennsylvania, 209 South 33rd Street, Philadelphia, Pennsylvania 19104, USA}
\affiliation{Department of Materials Science and Engineering, University of Pennsylvania, 3231 Walnut Street, Philadelphia, Pennsylvania 19104, USA}
\author{Xingting Gong}
\affiliation{Department of Physics and Astronomy, University of Pennsylvania, 209 South 33rd Street, Philadelphia, Pennsylvania 19104, USA}
\author{Euiyeon Jung}
\affiliation{Department of Materials Science and Engineering, University of Pennsylvania, 3231 Walnut Street, Philadelphia, Pennsylvania 19104, USA}
\author{Daniel M. Sussman}
\affiliation{Department of Physics and Astronomy, University of Pennsylvania, 209 South 33rd Street, Philadelphia, Pennsylvania 19104, USA}
\author{Shu Yang}
\affiliation{Department of Materials Science and Engineering, University of Pennsylvania, 3231 Walnut Street, Philadelphia, Pennsylvania 19104, USA}
\author{Randall D. Kamien}\email{kamien@upenn.edu}
\affiliation{Department of Physics and Astronomy, University of Pennsylvania, 209 South 33rd Street, Philadelphia, Pennsylvania 19104, USA}

\date{\today}

\begin{abstract}
In this paper we explore and develop a simple set of rules that apply to cutting, pasting, and folding honeycomb lattices.  We consider  {\sl origami}-like structures that are extinsically flat away from zero-dimensional sources of Gaussian curvature and one-dimensional sources of mean curvature, and our cutting and pasting rules maintain the intrinsic bond lengths on both the lattice and its dual lattice.  We find that a small set of rules is allowed providing a framework for exploring and building {\sl kirigami} -- folding, cutting, and pasting the edges of paper.

\end{abstract}

\renewcommand\vec[1]{\mathbf{#1}}
\newcommand\dd{\hbox{d}}
\newcommand{\Tr}{\mathop{\rm Tr}}
\renewcommand\vec[1]{\mathbf{#1}}

\maketitle

From pleating a skirt \cite{chai}, to wrapping a package \cite{package}, to folding an airplane \cite{rob1} or a robot \cite{rob2}, the art, technology, and mathematics of {\sl origami} explores the reach and breadth of what can be created from nearly unstretchable surfaces \cite{miura,maha}.  Flat-folded {\sl origami} is the folding of two-dimensional surface with zero Gaussian curvature everywhere \cite{torus}.  All the structure therefore arises from the extrinsic curvature of the sheet. Remarkably, the inverse problem -- how does one fold a target structure -- is algorithmically solved via a combinatorial procedure that creates the base of the final product \cite{lang}.  In this case, the paper is flat away from the sharp creases.  To exploit {\sl origami} for buildings, electronic circuits, robots, and metamaterials that are typically made of rigid plates \cite{chen}, flat regions joined only at sharp bends is a necessary design constraint. 
Though the sharp bends expose an exquisite interplay of bending and stretching in real materials \cite{witten}, we (and others) set that physics aside and consider only idealized, perfectly sharp folds in a non-shearable, non-stretchable medium.  With so much already understood, what new modalities are available to advance the state of `paper' art? Here we consider {\sl kirigami} of a rigid two-dimensional sheet with folds and {\it cuts} that remove topological discs from the original sheet.  In Fig. 1 we show a prototypical {\sl kirigami} design, inspired by the deep ideas of Sadoc, Rivier and Charvolin on phyllotaxis \cite{sadoc1,sadoc,sadoc2}.  By exploiting the connection between topology and intrinsic geometry we can add intrinsic curvature to sheets in a controlled manner \cite{sa,sb}, an effect which can be coupled with the extrinsic curvature techniques of origami. We develop a series of rules for lattice {\sl kirigami}, subject to some simplifying restrictions for simplicity of presentation and designability.

\begin{figure}[!ht] 
\centerline{\includegraphics[width=0.7\linewidth]{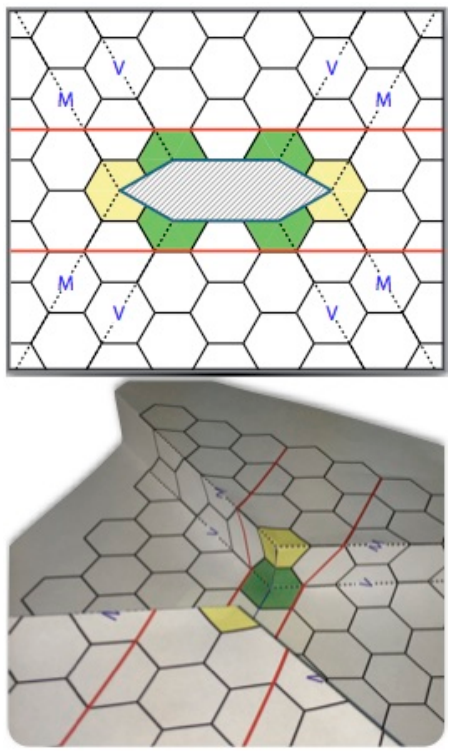}}
\caption{The essence of {\sl kirigami}.  Top: We remove the hatched region and make mountain (M) and valley (V) folds along the indicated lines.  Bottom: Final state with the edges of the cut identified by pasting.   }
\end{figure}

We develop our ideas on the honeycomb lattice: a natural starting point if we are considering fixed edge-length structures with, for instance, the minimum number of fixed-length struts per unit area \cite{Hales}, graphene and graphene-like materials \cite{qipark}, or self-assembled RNA networks \cite{gearyRNA}.  As we will show, enforcing a no-stretching condition on the bonds of the lattice strongly constrains the allowed cuts and folds, leading us to identify a small set of rules that can be used to build target structures. This cutting (and pasting) leaves us with co\"ordination number defects \cite{graphene} on both the honeycomb and its dual (triangular) lattice. 

There are natural restrictions to impose on the lattice {\sl kirigami}, both to respect the nature of the lattice, and also to simplify the development. (i) {\it We assume that our sheet cannot shear or stretch and can only be bent and cut along straight lines}.  Further, we insist that (ii) {\it edge lengths are preserved on the lattice and its dual}.  Throughout we will list assumptions and rules in italics preceeded by Roman or Arabic numerals, respectively.

To begin, consider Fig. 1: after making cuts and identifying edges we have a surface with all Gaussian curvature concentrated at the cone points corresponding to the corners of the cutout.  We can clearly see two 5-7 disclination pairs (pentagons and heptagons, respectively) and, by following the red lines, the dislocation/anti-dislocation pair they create \cite{hexatic1,hexatic2}.  We will refer to the triangular lattice of hexagon centers as $\tilde{\boldsymbol{\Lambda}}$ in the following and will label the defects accordingly so that, for instance, a five-fold disclination will be written $\tilde 5$.  Let  $\boldsymbol{\ell}$ (see Fig. 2) point from the dislocation to its corresponding partner in the anti-dislocation pair -- in Fig. 1 both $\tilde 7$'s -- while $\vec{b}$ is the Burgers vector of the dislocation.  The prototype in Fig. 1 is especially symmetric because $\boldsymbol{\ell}\!\perp\!\vec{b}$. Since the final configuration can be arrived at via a dislocation climb of a  dislocation/anti-dislocation pair we will call this geometry pure climb.  Similarly, if $\boldsymbol{\ell}||\vec{b}$ then we will refer to that geometry as pure glide (see Fig. 3); general configurations will have both glide and climb.  It is important to note that  from an intrinsic point of view the bond lengths are kept fixed and all the polygons are regular, but the extrinsic geometry is naturally distorted as the structure moves to three dimensions.  Finally, we note that there is a degeneracy in the folded structure.  Each plateau can individually `pop up' or `pop down.'  This extra degree of freedom should prove useful in the targeted design of structures \cite{tbp}. In the supplemental material we provide cutting and folding templates for the home scientist -- the template for Fig. 1 is in Fig. S1. 

We can also preserve intrinsic bond lengths by working on a second triangular lattice, ${\boldsymbol{\Lambda}}$, the Bravais lattice of the honeycomb. The honeycomb is not itself a Bravais lattice, but instead a lattice with a basis.  It is necessary to interpret topological defects from the point of view of the underlying Bravais lattice, excitations of which are the Goldstone modes \cite{peanut}.  The vectors $\vec{e}_1=\left[0,1\right]$ and $\vec{e}_2=\left[\sqrt{3}/2,1/2\right]$ are the basis vectors for $\tilde{\boldsymbol{\Lambda}}$ and ${\boldsymbol{\Lambda}}$.  The two Bravais lattices are offset by the displacement $\vec{d}=(\vec{e}_1-2 \vec{e}_2)/3$.  Each Bravais lattice site on the honeycomb has a two-vertex basis, one at the lattice site and the other at a displacement of $\boldsymbol{\delta}=\left[-\sqrt{3}/3,0\right]$, up to rotation by $2\pi/3$.  As shown in Fig. 2 (and Fig. S2), a 2-4 defect pair in $\boldsymbol{\Lambda}$ will appear as a pair of neighboring points on the honeycomb lattice with 2-fold and 4-fold co\"{o}rdination.  Note that when making the cut in this case assumption (ii) makes it necessary to pairwise identify {\it both} vertices in the basis so that we cut out a $2\pi/3$  wedge instead of $\pi/3$ wedge as in the defect pair in $\tilde{\boldsymbol{\Lambda}}$ on the right of Fig. 2.  From this point of view the 2-4 is a topological misnomer, though one which we will continue to use for clarity.  It is actually a 5-7 pair  on $\boldsymbol{\Lambda}$ following the same rules that apply to $\tilde{\boldsymbol{\Lambda}}$.  Finally, because the 2-4  and $\tilde 5$-$\tilde 7$ defects are separated by a vector that is not on either lattice, but rather a lattice vector plus $\bf d$, we will refer to this configuration as a {\it partial} climb -- a non-integer number of intervening sites must be removed to form the structure in Fig. 2. Thus our first rule, (1) 
{\it the vector $\boldsymbol{\ell}$ between two disclinations -- cut corners of which are located on either ${\boldsymbol{\Lambda}}$ or $\tilde{\boldsymbol{\Lambda}}$ -- can be composed of glide and climb components}.

\begin{figure}[t] 
\centerline{\includegraphics[width=0.9\linewidth]{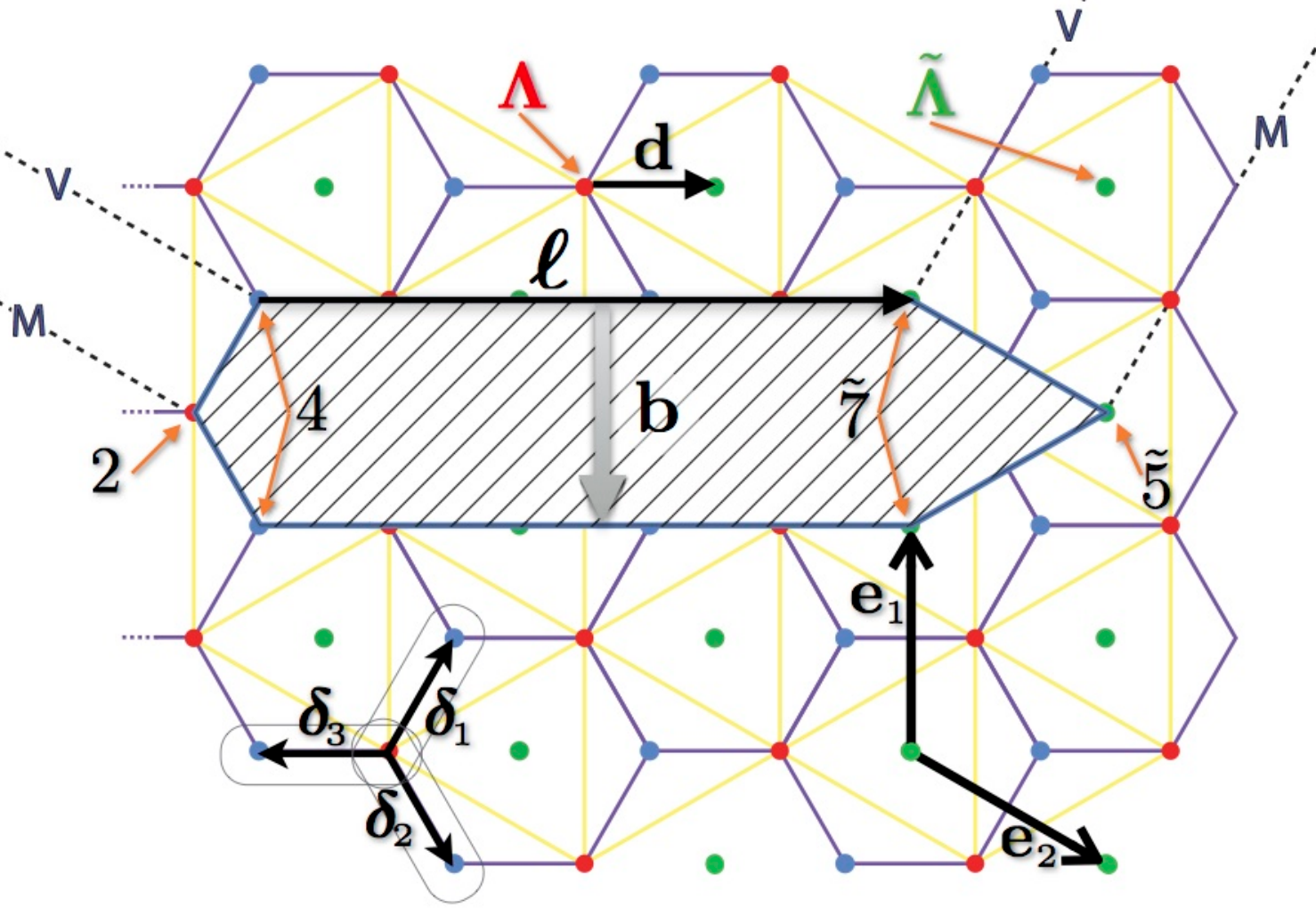}}
\caption{The two lattices $\boldsymbol{\Lambda}$ (red dots) and $\tilde{\boldsymbol{\Lambda}}$ (green dots) offset by ${\bf d}$.  Yellow edges are on $\boldsymbol{\Lambda}$.  We show the basis vectors of the lattices$\{{\bf e}_i\}$ and the unit cells $\{\boldsymbol{\delta}_i\}$.  A `2-4' pair on the honeycomb is a standard 5-7 pair on $\boldsymbol{\Lambda}$.  The cut is absorbed by the $\tilde{5}$-$\tilde{7}$ pair on $\tilde{\boldsymbol{\Lambda}}$ creating a partial climb.  The plateaus of the 2-4 and $\tilde{5}$-$\tilde{7}$ pairs are {\it different} heights upon folding.}
\end{figure}

In Fig. 1  the sidewalls of the plateaus are vertical so their height is the lattice constant of $\tilde{\boldsymbol{\Lambda}}$.   (iii) {\it We will restrict ourselves to vertical sidewalls from this point on}. This requires that (2) {\it folds terminating at a corner of a disclination's excised triangle must be perpendicular to the cut edges.}  It follows that the angle of each plateau corner is the supplement of the excised angle so the folds are also commensurate with the lattice.

Can cuts other than the 2-4 and $\tilde 5$-$\tilde 7$ be made that still preserve the both lattices? No: around a point of $N$-fold symmetry, only cuts and rejoins that are multiples of $2\pi/N$ bring different vertices into coincidence and  preserve all lattice distances.  The honeycomb lattice has points of $2$-, $3$-, and $6$-fold symmetry, but the $2$-folds are not suitable sites. They occur at the mid-edges of both $\boldsymbol{\Lambda}$ and $\tilde{\boldsymbol{\Lambda}}$, so the formation of a $\pi$ cone point leaves both lattices with dangling half-edges. The only points around which we can make the cuts are thus on the vertices of $\boldsymbol{\Lambda}$, $\boldsymbol{\Lambda}+{\bf d}$, and $\tilde{\boldsymbol{\Lambda}}$. It follows that (3) {\it the 2-4 and $\tilde 5$-$\tilde 7$ pairs are the basic building blocks of hexagonal lattice {\sl kirigami}.} Other motifs such as $\tilde 4$-$\tilde 8$ can be made by combining two $\tilde 5$-$\tilde 7$ pairs in the appropriate way.

The plateau created by the $\tilde 5$-$\tilde 7$ is $\sqrt{3}$ taller than the plateau of the 2-4 as the geometry of Fig. 2 dictates, so these dislocations cannot be mixed around the boundary of a shared plateau. Though cuts can cross, folds cannot without creating extra cuts in the original sheet -- a dihedral angle in a vertical wall must be accompanied by a new dislocation.  Because of the height differences we have (4) {\it plateaus must be surrounded by 2-4 pairs (triangles) or $\tilde 5$-$\tilde 7$ pairs (hexagons), but they cannot be mixed.}
\begin{figure}[t] 
\centerline{\includegraphics[width=0.9\linewidth]{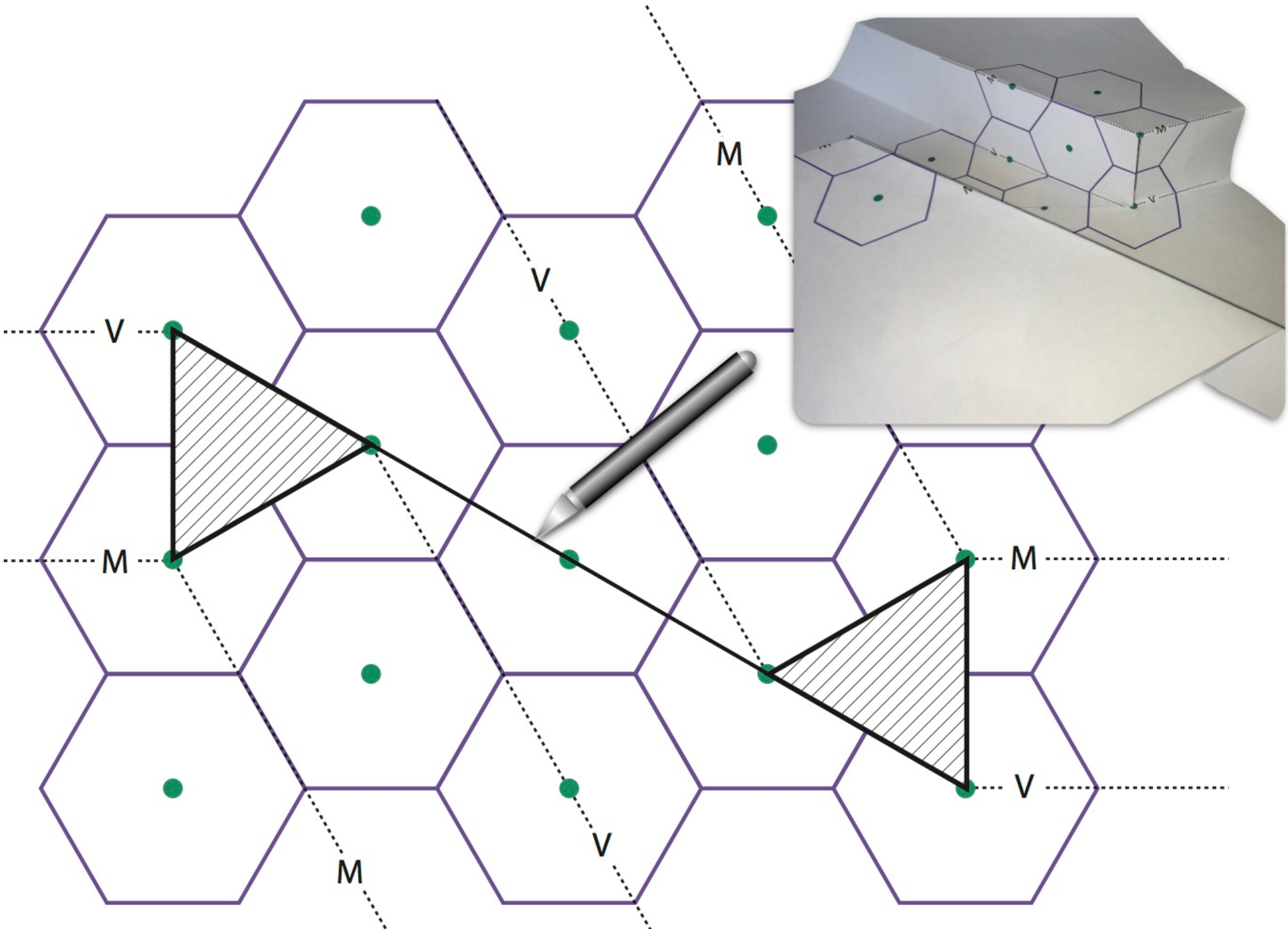}}
\caption{The glide cut.  We remove the hatched region and make mountain (M) and valley (V) folds along the indicated lines.  The inset shows the folded geometry.  Note the cut along the line indicated by the knife.}
\end{figure}
Finally, note that the 2-4 and $\tilde 5$-$\tilde 7$ motifs of Fig. 2 both result in a strip of equal width being excised, which is necessary for them to coexist at each end of one single {\sl kirigami} cut. A further requirement is that the disclination/anti-disclination pair are facing each other, without which the two sides of the vacant strip would mismatch when pasted together. This geometric argument can be summed up by the usual rule for topological defects, (4) {\it dislocations can cancel when their Burgers vectors sum to zero}. It follows that we can also bring cuts together along three directions to form a triangular `plateau'. If the tips of the excised triangles of dislocations all coincide then the plateau is of zero height, and the dislocations cancel within the plane.

The examples shown so far only consider climb dislocation pairs. {\sl Kirigami} also allows for pure glide, with  $\boldsymbol{\ell}||\vec{b}$, in which no extra material is added or removed in the dislocation itself.  Consider two $\tilde 5$-$\tilde 7$ defects as in Fig. 3 (Fig. S3).  We remove two equilateral triangles separated by a cut perpendicular to the disclination dipole direction, along the Burgers vector.  Bringing the edges of the triangles together slides one edge of the cut along the other.  From the point of view of actuators, this mode of popping into the third dimension can have all the mechanics built onto the paper -- a ratchet-and-pawl or rack-and-pinion could be manufactured into the initial sheet.  By putting together the geometries in Figs. 1 and 3, we can add climb to a glide by extending either triangular cutout along a climb direction.  We can replace a $\tilde 5$-$\tilde 7$ in Fig. 3 with a 2-4 pair, but only by adding a partial climb -- a 2-4 and $\tilde 5$-$\tilde 7$ pair cannot be a pure glide.

Looking at the cutting patterns for dislocation pairs shows the possibilities afforded by {\it kirigami} that pure {\it origami} lacks. There are clever techniques within origami for forming wedges and pleats to `remove' material by tucking it behind the visible surface, creating dipoles of Gaussian curvature similar to our dislocations. However, a consideration of, for example, the glide dislocations of Fig. 3 indicates that the structures created so simply with {\it kirigami} techniques require dramatically plicated correspondences in pure {\it origami}.  This unnecessary  complexity involves not only many additional folds but large swathes of `wasted' triple thickness paper. In this way using traditional {\sl origami} to achieve a sixon is reminiscent of the historical introduction of epicycles into the supposedly circular orbits of planets in order to correct for the differences between theory and observation, instead of changing paradigm to elliptic orbits.  From a design perspective the simplicity of {\sl kirigami} is thus seen in contrast to the complex folding sequence of pure {\sl origami} needed to achieve a target structure.

Up until this point we have considered cutout regions for which the identification of edges is set by the geometry -- long edges join with long edges.  However, we may also consider a completely symmetric situation.  Consider the cuts and folds in Fig. 4a. (Fig. S4).    If we were to cut out the hatched region we would create a vacancy in $\tilde{\boldsymbol{\Lambda}}$.  There are three degenerate ways to fold the up/down structure formed by identifying any two pairs of parallel edges.  Once we break the degeneracy (Fig. 4b) we can fold one pair and then bring the two dislocations together (Fig. 4c)  to form the final state that is independent of our choice of initial folds (Fig. 4d).   The resulting state is reminiscent of the instability that leads to martensites and tweeds in crystals  when a particular crystal habit can be distorted along different paths to a new crystal habit -- for example the three different Bain strains that turn the FCC lattice into the BCC lattice \cite{bhatt}.

\begin{figure}[t] 
\centerline{\includegraphics[width=0.9\linewidth]{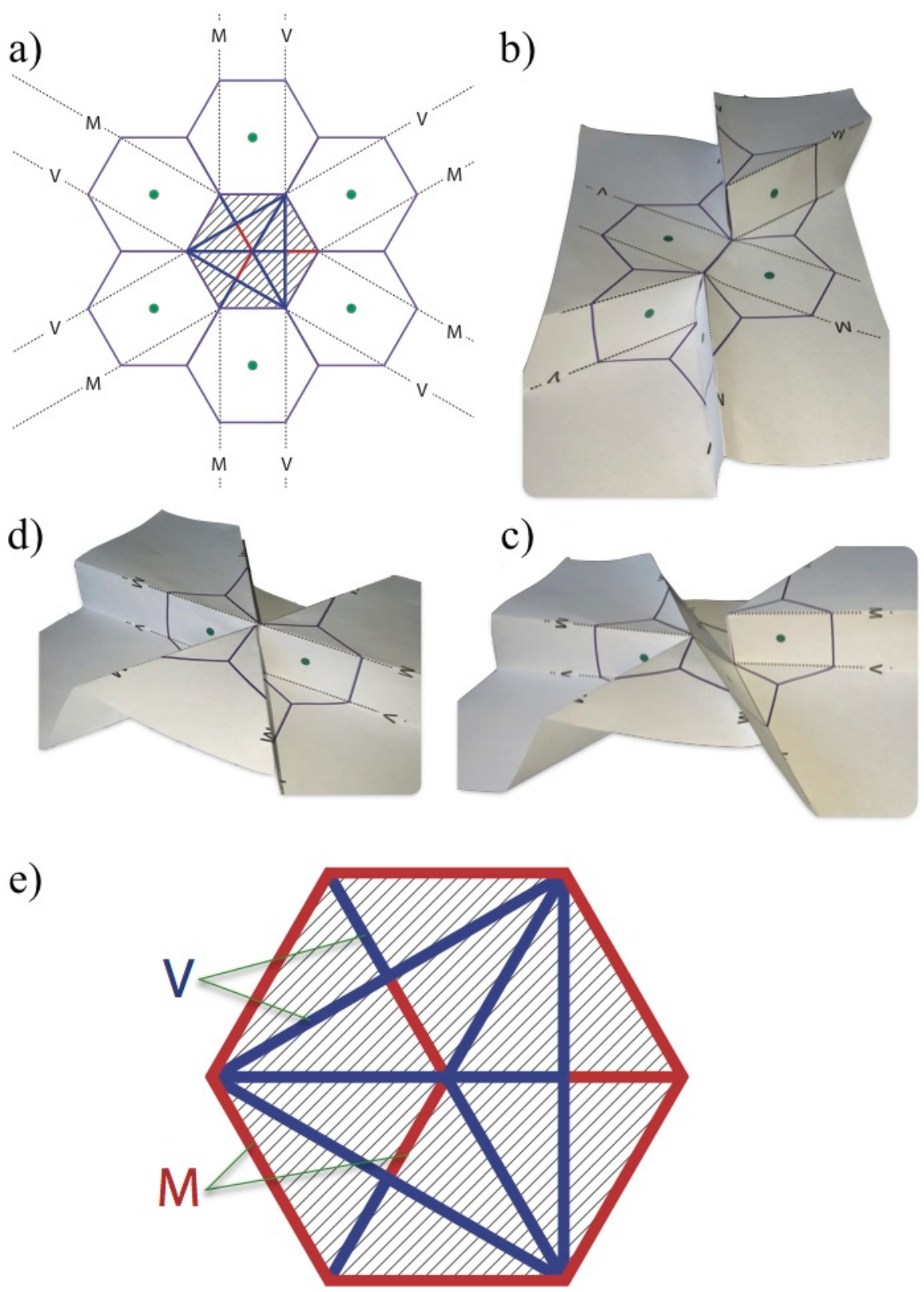}}
\caption{The `sixon.'  a) Basic template. b) This structure is degenerate and can be split three different ways into matching 2-4 pairs with pop up and pop down configurations. c) An intermediate state connecting b) to the final state d).  The structure in d) can be filled in with the hexagon in e) with mountain and valley folds  made along the red and blue lines, respectively.}
\end{figure}

We note that the three-fold structure in Fig. 4d, which we dub the `sixon,' can also be formed by pure {\sl origami} with the set of mountain and valley folds illustrated in Fig. 4a. This construction similarly brings all three vertices together and effects a vacancy by folding the paper of the `excised' hexagon underneath the surface, although it is an open question whether the configuration is rigid-foldable. If not, previous results suggest that this pattern could be made rigidly foldable \cite{demaine}, but only at the cost of adding extra (and probably sub-lattice) folds. On the other hand, by excising the interior hexagon both the detached hexagon of Fig. 4e and the {\sl kirigami} of Fig. 4d can be rigidly folded individually, and the excised hexagon reattached if desired within the cut-and-paste rubric of {\it kirigami}.

 \begin{figure}[t] 
\centerline{\includegraphics[width=0.7\linewidth]{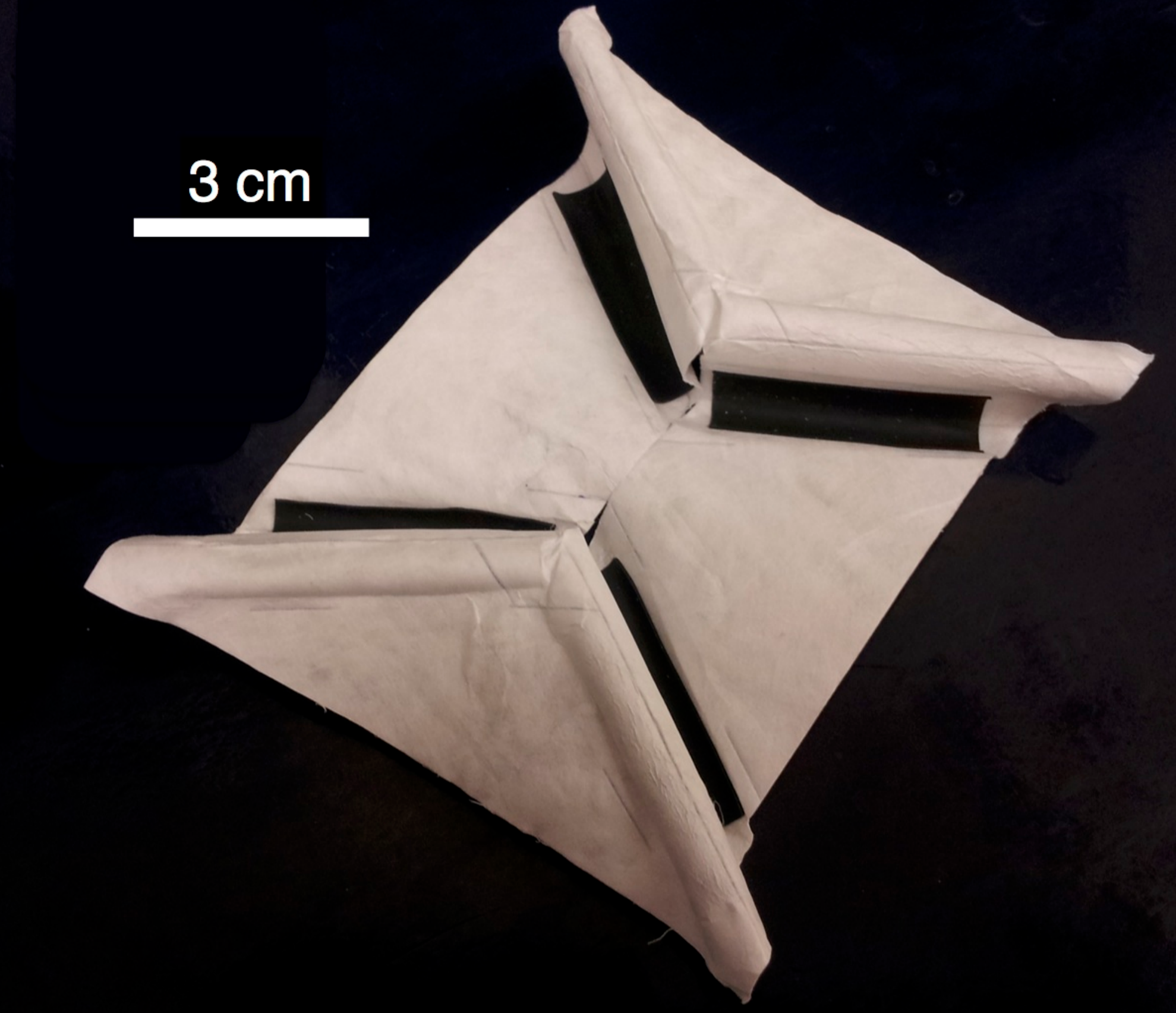}}
\caption{Tyvek and polyolefin {\sl kirigami}.  The black strips are the polyolefin.  The cuts outlined in Fig. 1 self seal with no intervention or glue (see supplemental movie).}
\end{figure}
To complement the paper models in Figs. 1, 3, and 4b-d, we have used Tyvek  (nonwoven, Spunbonded Olefin, Type 10) as our two-dimensional material because it is stronger and more tear-resistant than standard paper.  We cut the pattern in Fig. 1 onto the Tyvek and bonded heat-shrinkable polyolefin (SPC Technology) along the mountain and valley folds by hot pressing at 120$^\circ$C for 20 minutes.  To prevent shrinking during bonding, the assembly was kept pressed until cooled to room temperature.  Subsequent unpressed baking at 95$^\circ$C for one minute curls the polyolefin strips and folds the Tyvek into the target structure shown in Fig. 5.  This autonomous self-assembly of three-dimensional target structures demonstrates the robustness of our {\sl kirigami} rules.

We may create ever more embellished structures by relaxing some of the rules.  For instance, the identified edges of a cut need not be straight lines as long as they are separated by the Burgers vector.  One could use this, for instance, to make serrations along a glide cut to lock the structure rigidly.  We can also consider structures that go off lattice to create non-vertical sidewalls, either overhanging or reclining. We may also mix the rules on the original lattices with rules on larger sublattices obtained by rotating the original lattices to create a {\it coincidence lattice} as in moir\'e patterns \cite{moire}.  These and other extensions will be considered elsewhere \cite{tbp}.

In summary, we have developed a small set of rules that minimally distort both an underyling honeycomb lattice of bonds and its dual to achieve localized Gaussian curvature and three-dimensional structure.  These rules can be understood in terms of the standard topological theory of dislocations and disclinations with the addition of intrinsic geometry to prevent stretching of an underlying, rigid material.  Finally, we have shown that the interplay of cuts and folds limits the allowed interactions of defects on the sheet.   Future work might consider these sorts of constructions on pre-curved  \cite{scars,scarsc,optics} or pre-swelled \cite{efi,comment,orimass,brain} sheets, in particular riffs on the regular tesselations of the hyperbolic plane as popularized by M.C. Escher \cite{escher}. We believe that these rules and ideas, coupled with clever materials design \cite{chen,qipark} will lead to new and useful ideas, modalities, and devices.

We acknowledge stimulating discussions with B.G. Chen, D.R. Nelson, J.-F. Sadoc, and C.D. Santangelo.
The authors acknowledge support from NSF EFRI-ODISSEI Grant EFRI 13-31583.  D.M.S. was supported by a fellowship from the American Philosophical Society. This work was partially supported by a
Simons Investigator grant from the Simons Foundation to
R.D.K.


\begin{thebibliography}{10}

\bibitem{chai} W.T.M. Irvine, V. Vitelli, and P.M. Chaikin, Nature {\bf 468}, 947 (2010).

\bibitem{package} C. Py, P. Reverdy, L. Doppler, J. Bico, B. Roman, and C.N. Baroud, Phys. Rev. Lett. {\bf 98}, 156103 (2007).

\bibitem{rob1} E. Hawkes, B. An, N.M. Benbernou, H. Tanaka, S. Kim, E.D. Demaine, D. Rus, and R.J. Wood, Proc. Natl. Acad. Sci. {\bf 107}, 12441 (2010).

\bibitem{rob2} S. Fulton, M. Tolley, E. Demaine, D. Rus, and R. Wood, Science {\bf 345}, 644 (2014).



\bibitem{miura} K. Miura, Proceedings of the 31st Congress of the International Astronautical Federation, IAF-80-A 31, (American Institute for Aeronautics and Astronautics, New York, 1980), pp. 1-10.

\bibitem{maha} L. Mahadevan and S. Rica, Science {\bf 307}, 1740 (2005).

\bibitem{torus} V. Borrelli, S. Jabrane, F. Lazarus, and B. Thibert, Proc. Natl. Acad. Sci., {\bf 109}, 7218 (2012).

\bibitem{lang} R.J. Lang, in {\it Proceedings of the Twelfth Annual Symposium on Computational Geometry} (ACM, New York, 1996), pp. 98--105.




\bibitem{chen} S. Waitukaitis, R. Menaut, B.G. Chen, and M. van Hecke, arXiv:1408.1607 (2014).


\bibitem{witten} T.A. Witten, Rev. Mod. Phys. {\bf 79}, 643 (2007).




\bibitem{sadoc1} J.-F. Sadoc, N. Rivier, and J. Charvolin, Acta Cryst. A{\bf68}, 470 (2012).
\bibitem{sadoc} J. Charvolin and J.-F. Sadoc, Biophys. Rev. and Lett. {\bf 6}, 13 (2011).
\bibitem{sadoc2} J.-F. Sadoc, J. Charvolin, and N. Rivier, J. Phys. A: Math. Theor. {\bf 46}, 295202 (2013).

\bibitem{sa} N. Bende, R.C. Hayward, and C.D. Santangelo, Soft Matter {\bf 10}, 6382 (2014).

\bibitem{sb} R. Kupferman, M. Moshe, and J.P. Solomon, arXiv:1306.1624 (2013).

\bibitem{Hales} T.C. Hales, Discrete Comput. Geom. {\bf 25}, 1 (2001).


\bibitem{qipark} Z. Gi, D.K. Campbell, and H.S. Park, arXiv:1407.8113 (2014).

\bibitem{gearyRNA} C. Geary, P.W.K. Rothemund, and E.S. Andersen, Science {\bf 345}, 799 (2014).


\bibitem{graphene} A. Kumar, M. Wilson, and M.F. Thorpe, J. Phys. Condens. Matter {\bf 24}, 485003 (2012).



\bibitem{hexatic1} B.I. Halperin and D.R. Nelson, Phys. Rev. Lett. {\bf 41}, 121 (1978).

\bibitem{hexatic2}
D.R. Nelson and B.I. Halperin, Phys. Rev. B {\bf 19}, 2457 (1979).

\bibitem{tbp} T. Castle, Y. Cho, X. Gong, D.M. Sussman, and R.D. Kamien, {\it in preparation} (2014).

\bibitem{peanut} S.J. Gerbode, U. Agarwal, D.C. Ong, C.M. Liddell, F. Escobedo, and I. Cohen, Phys. Rev. Lett. {\bf 105}, 078301 (2010).



\bibitem{bhatt} K. Bhattacharya, S. Conti, G. Zanzotto, and J. Zimmer, Nature {\bf 428}, 55 (2004).



\bibitem{demaine} E.D. Demaine, M.L. Demaine, V. Hart, G.N. Price, and T. Tachi, Graphs and Combinatorics {\bf 27}, 377 (2011).




\bibitem{moire} R.D. Kamien and D.R. Nelson, Phys. Rev. Lett. {\bf 74}, 2499 (1995).





\bibitem{optics} C.D. Santangelo, V. Vitelli, R.D. Kamien, and D.R. Nelson, Phys. Rev. Lett. {\bf 99}, 017801 (2007).


\bibitem{scars} A.R. Bausch, M.J. Bowick, A. Cacciuto, A.D. Dinsmore, M.F. Hsu, D.R. Nelson, M.G. Nikolaides, A. Travesset, and D.A. Weitz, Science {\bf 299}, 1716 (2003).

\bibitem{scarsc} R.D. Kamien, Science {\bf 299}, 1671 (2003).


\bibitem{efi} Y. Klein, E. Efrati, and E. Sharon, Science {\bf 315}, 1116 (2007).




\bibitem{comment} R.D. Kamien, Science {\bf 315}, 1083 (2007).


\bibitem{orimass} J. Kim, J.A. Hanna, M. Byun, C.D. Santangelo, and R.C. Hayward, Science {\bf 335}, 1201 (2012).

\bibitem{brain} T. Tallinen, J.Y. Chung, J.S. Biggins, and L. Mahadevan, Proc. Natl. Acad. Sci. {\it Early Edition} doi.10.1073/pnas.1406015111 (2014).



\bibitem{escher} M. C. Escher, Circle Limit III \& Circle Limit IV; see also C.D. Modes and R.D. Kamien, Phys. Rev. E {\bf 77}, 041125 (2008).








\end{thebibliography}
\end{document}